\documentclass[10pt,conference]{IEEEtran}
\IEEEoverridecommandlockouts

\usepackage{tcolorbox}
\usepackage{listings}
\usepackage{url}

\begin{document}

\title{Simulated Interactive Debugging}

% \author{\IEEEauthorblockN{Anonymous Authors}}

\author{\IEEEauthorblockN{Yannic Noller\IEEEauthorrefmark{1},
Erick Chandra\IEEEauthorrefmark{2}, Srinidhi Chandrashekar\IEEEauthorrefmark{2},
Kenny Choo\IEEEauthorrefmark{2}, Cyrille Jegourel\IEEEauthorrefmark{2},\\ Oka Kurniawan\IEEEauthorrefmark{2}, Christopher M. Poskitt\IEEEauthorrefmark{3}}
\IEEEauthorblockA{\IEEEauthorrefmark{1}\textit{Ruhr University Bochum}, Germany 
\IEEEauthorrefmark{2}\textit{Singapore University of Technology and Design}, Singapore\\
\IEEEauthorrefmark{3}\textit{Singapore Management University}, Singapore\\
\IEEEauthorrefmark{1}yannic.noller@acm.org,
% \IEEEauthorrefmark{2}\{erick\_chandra, srinidhi\_chandrasheka, kenny\_choo, cyrille\_jegourel,\\ oka\_kurniawan\}@sutd.edu.sg,
\IEEEauthorrefmark{2}\{name\_surname\}@sutd.edu.sg,
\IEEEauthorrefmark{3}cposkitt@smu.edu.sg}}

\maketitle

\begin{abstract}
Debugging software, i.e., the localization of faults and their repair, is a key activity in software engineering. Therefore, effective and efficient debugging is one of the core skills a software engineer must develop. However, the teaching of debugging techniques is usually very limited or only taught in indirect ways, e.g., during software projects. As a result, most Computer Science (CS) students learn debugging only in an ad-hoc and unstructured way.
In this work, we present our approach called \textit{Simulated Interactive Debugging} that interactively guides students along the debugging process. The guidance aims to empower the students to repair their solutions and have a proper \textit{learning} experience. We envision that such guided debugging techniques can be integrated into programming courses early in the CS education curriculum.
We developed a prototypical implementation using traditional fault localization techniques and large language models. Students can use features like the automated setting of breakpoints or an interactive chatbot. We designed and executed a small-scale, controlled experiment with eight undergraduate CS students. Based on the responses, we conclude that the participants liked the systematic guidance. They rated the automated setting of breakpoints as most effective, followed by the interactive debugging and chatting, and the breakpoint explanations.
In future, we will extend our concept and implementation, and perform more intensive user studies.
\end{abstract}


\section{Introduction}
% Context
Despite the paradigm shift towards Artificial Intelligence (AI)-assisted software development, particularly with the emergence of Large Language Models (LLMs), there remains a high demand for qualified software engineers capable of developing reliable, high-quality software.
With more auto-generated code, it will become even more critical that junior software developers can effectively debug software and solve bugs, even in code they have not written themselves.
%
% Open Challenge(s)
Radermacher et al.~\cite{radermacher_icse2014} explored knowledge deficiencies of graduate students from an industry perspective. One of the most frequently mentioned issues with software tools has been version control systems and debuggers. While we have seen the integration of version control systems like Git in the Computer Science (CS) and Software Engineering (SE) curriculum, e.g., with the usage of GitHub Classroom, we have not seen much innovation regarding debugging education. 
Michaeli and Romeike~\cite{Michaeli2019} noted that there are only a few studies investigating the ``explicit teaching of debugging'' and students are often left alone to learn debugging ``the hard way''.
Our experience confirms this observation: most students still learn (interactive) debugging of code in an ad-hoc and unstructured way by trial and error. Using ``print/log'' statements is often the only extent of students' experience in debugging. They do not receive proper guidance in efficient program comprehension and how to debug programs effectively.
%
% Related Work
Current CS/SE research efforts focus on general programming education, which helps to provide scalable alternatives to mentor and guide a rising number of students. Automated Program Repair (APR) techniques~\cite{clara_pldi2018,refactory_ase2019,pydex_oopsla2024} can help to produce patches for incorrect submissions, which then can be used as a basis for personalized feedback, or even automated grading~\cite{fan_issta2023}. In fact, we already have seen the application of such techniques in the CS/SE curriculum focusing on repairing students' solutions and providing feedback~\cite{codeaid_chi2024, fan2024softwareengineeringeducationalexperience}. However, the existing related work in CS education does not address actual debugging.
%
% Our research objective
% \begin{tcolorbox}[boxrule=1pt,left=1pt,right=1pt,top=1pt,bottom=1pt]
Therefore, our research objective is to \textit{understand} students' needs in learning to debug and provide \textit{automated} means to \textit{guide} them effectively along the debugging process.
% \end{tcolorbox}
%
% As part of this objective, we strive to seek answers to questions like "How can we guide the debugging process using artifacts generated from APR/SE techniques?" and "How can LLMs be integrated into the debugging process for the benefit of the student?"

% Our proposal to solve the problem.
As a first step to achieve our research objective, we propose \textit{Simulated Interactive Debugging}. The essence of this concept is to create a simulation of an interactive debugging experience; ``simulated'' because the solution is known in the education context due to given reference implementations and auto-generated artifacts like fault locations and patches. Our goal is to provide a controlled but supportive environment, in which the students debug the problem on their own. Similar to a human tutor, we would not directly reveal the solution and instead help the student to understand and fix the issue. Our vision of such an assisted debugger requires integration into coding tools like Integrated Development Environments (IDEs) to actively guide students.
Based on APR, fault localization, and LLMs, we can show potential problematic locations, explain the issues at these locations, set breakpoints, identify relevant variable values, and provide repair hints.

% “Design science methodology is focused on building and evaluating artifacts that are designed to address a problem. The overall aim is to derive generalized knowledge about problems and related solutions. Design science methodology, as a problem-solving process, was introduced by Hevner et al. [106]. It requires the creation of an artifact for a specific problem. The artifact should be innovative and effective. Furthermore, it needs to be evaluated using rigorous research methods. Design science comes from information systems. Wieringa [266] helped bridge the gap from information systems to software engineering. Design science in software engineering is further explored by Runeson et al. [213] and Engström et al. [70].”
% “106.Hevner, A.R., March, S.T., Park, J., Ram, S.: Design science in information systems research. MIS Q. 28(1), 75–105 (2004). https://doi.org/10.2307/25148625” 
% “266.Wieringa, R.J.: Design Science Methodology for Information Systems and Software Engineering. Springer, Berlin (2014). https://doi.org/10.1007/978-3-662-43839-8”

% Current state of our research
As part of a design science methodology~\cite{hevner2005_designscience,wieringa2014_designscience}, we developed a prototype as a VS Code IDE extension and conducted a pilot user study with eight undergraduate CS students. The results indicate that students have little experience with (interactive) debugging but enjoy the guidance, notably the automatic breakpoint setting and the interaction with the chatbot.
%
% Summary
In summary, our core contributions are:
\begin{itemize}
    \item the concept of \textit{Simulated Interactive Debugging} to guide CS students along the interactive debugging process
    \item the implementation of this concept as an \textit{IDE plugin}
    \item a \textit{pilot user study} to show the practicality and usability of our concept and its implementation
\end{itemize}
%
% Future Work and Vision.
% As a next step, we will use the feedback we have collected to improve our concept and its implementation and add more features.
% Our vision of simulated interactive debugging will complement existing programming education teaching and help students efficiently learn the relevant debugging and program comprehension skills.

\section{Related Work}

\paragraph{Intelligent Tutoring Systems}
% General Issue
Due to the rise in the number of CS students~\cite{masters2011}, educators need help to cope with the number of assignment submissions. % and the needed feedback. 
% Automated Program Repair
APR techniques are currently being explored to supplement the missing human tutor, e.g., by realizing so-called Intelligent Tutoring Systems (ITS).
% Traditional APR for ITS
Early works in this domain attempt to repair student submissions to generate feedback~\cite{clara_pldi2018,refactory_ase2019} and automatically grade programming assignments~\cite{fan_issta2023,birillo_sigcse2022}.
% LLM-based APR for ITS
More recently, LLM-based approaches have been explored to generate personalized feedback and hints for programming assignments~\cite{pydex_oopsla2024,codehelp_calling2023,hou_ls2024,zhao2024peer,kumar_llm4code2024}. Several experiences have been reported where LLM-based programming assistants have been deployed in the classroom context~\cite{codeaid_chi2024,fan2024softwareengineeringeducationalexperience,abolnejadian_chiea2024,kurniawan_tale2023}.
LLMs also can be combined with traditional techniques like static analysis to provide incremental hints for programming tasks~\cite{birillo_koli2024}.
Moreover, existing works~\cite{koutcheme2024benchmarking, phung_icer2023} benchmarked various models regarding their capabilities to support programming education.
Overall, the existing works focus on repairing programming assignments and generating feedback; they do not address the teaching of interactive debugging.

\paragraph{Debugging Education}
% Program debugging is an area that has long been researched and still is an active research area.
To get an overview of the teaching of debugging, we refer to McCauley et al.~\cite{McCauley01062008}, who conducted a systematic literature review.
Michaeli and Romeike~\cite{Michaeli2019} recently explored the influence of teaching systematic debugging concepts with an intervention study concluding that explicitly teaching debugging skills positively affects debugging self-efficacy.
Recent LLM-inspired, debugging-related research focuses on Socratic questioning~\cite{alhossami_sigcse2024} and generating so-called debugging quizzes~\cite{phung_icer2023,padurean2024bugspotter} that ask students to reason about a given buggy program and design bug-revealing test cases.
%
% While the reported experiences and the existing approaches and tools can help students build general debugging skills, to the best of our knowledge, no approach attempts to guide students through the interactive debugging process.
While the reported approaches and tools can help students build general debugging skills, to the best of our knowledge, no approach guides through the interactive debugging process.

\paragraph{AI in CS Education}
Most recently, with the emergence of LLMs, AI has arrived in the domain of CS education. Shein~\cite{shein_cacm2024} argues that while we still need to teach students the fundamentals of programming, such teaching will become more ``alive'' through AI, e.g., via chatbots.
% Traditionally, Learning Management Systems (LMS) help educators manage the load of submissions, provide teaching material, and auto-grade assignments. Systems like Artemis~\cite{artemis_sigcse2018} even provide an interactive learning experience and individual feedback for various exercise types. Recently, Iris~\cite{bassner_iticse2024} was integrated into Artemis, providing an AI-based, virtual tutor to support students in programming assignments.
Denny et al.~\cite{denny_iticse2024} investigated desirable characteristics for AI teaching assistants and concluded that students enjoy AI-based learning support and actually preferred tailored scaffolding instead of mere responses and the revealing of solutions.
Kazemitabaar et al.~\cite{kazemitabaar2024exploring} explored the design space for such AI tools and highlighted the importance of step-by-step guidance and an interactive dialog with the AI.
Following the insights of these works, our proposed concept for Simulated Interactive Debugging guides the student in understanding the problem and identifying the programming faults instead of just providing feedback on how to fix it. Further, we also offer an AI-based chatbot with which the students can engage in a controlled environment.

\section{Concept: Simulated Interactive Debugging}
\label{sec:concept}
To tackle the shortcomings of CS education in program debugging, we propose the following concept. Using technology from automated SE, in particular from APR, we can identify the issue in a student's buggy program and can generate a corresponding solution. Note that such a solution can structurally and syntactically differ from the reference implementation provided by a lecturer. Now, knowing the fault and where to fix it, we can use this knowledge to guide the student not only in fixing the issue, but step-by-step in understanding the defect, identifying the potential fix locations, and eventually fixing the problem---similar to how a human tutor would guide the student along the debugging process. We designed and developed an AI-assisted debugger that offers students a \textit{simulated} interactive debugging experience. Students receive help in running tests, using the interactive debugger in identifying the underlying fault in the problem, and in fixing the actual problem.
As Birillo et al.~\cite{birillo_ide2024} have argued, combining programming education with the use of IDEs helps familiarize students with industrial technologies; hence we envision our concept being deployed within IDEs.

\paragraph{The Role of LLMs}
This concept fits into a larger research endeavor of ours, where we aim to develop an active learning environment integrating APR through LLMs and other SE tools with interactive debugging. This concept is designed to provide interactive guidance to students, thereby enhancing their learning experience in programming. Furthermore, our concept exposes students to AI tools, in particular LLMs, in a guided and controlled way. LLMs can be deployed in various roles in this concept, e.g., to generate hints and explanations in natural language, as well as core technology for an interactive chatbot.
Based on the generated artifacts from techniques like spectrum-based fault localization (SBFL)~\cite{fl_survey_tse2016}, static code analysis, symbolic execution, code synthesis and LLM prompting, we can set breakpoints automatically, identify relevant variable values, and highlight them for the students. Such a process can also be supported by Socratic questioning~\cite{alhossami_sigcse2024}.

\paragraph{Workflow}
%
% Figure~\ref{fig:workflow} shows an exemplary workflow that implements the idea of simulated interactive debugging. 
%
The concept has two types of users: the students and the lecturers, although the main focus is on the students.
Both user groups provide inputs and receive feedback. The lecturer provides the \textit{programming assignment}, the \textit{reference implementation}, and a test suite consisting of \textit{private} and \textit{public test cases}. The student provides a \textit{buggy solution} and (optionally) a set of custom test cases.
There are three different use cases: (1) the primary use case is the assisted debugging, (2) the test assessment for \textit{lecturers}, and (3) the test assessment for \textit{custom tests} from the student.
The test assessment for \textit{lecturers} checks whether the given test suite is strong enough to catch errors and perform effective fault localization, e.g., via mutation testing.
% This can be realized via mutation testing, and additional tests could be generated, e.g., via evolutionary test generation~\cite{evosuite_fse2011}.
% By offering additional features for the lecturer, our goal is also to increase confidence in the overall system, which is crucial for its successful deployment in programming courses.
The test assessment for \textit{custom tests} from the student works with the idea that when the student submits custom tests, these can be executed against the reference implementation from the lecturer.
Failing test cases can be flagged and corrected; further, we can propose new tests, e.g., to explore boundary cases or to increase coverage.
% If they fail on the reference implementation, they are flagged and can also be corrected (because the correct output for the inputs of such test cases could be extracted via the reference implementation). To help students, we can also propose new test cases to them, e.g., to explore certain boundary cases or to increase overall code coverage.

% \paragraph{Deployment}
% Birillo et al.~\cite{birillo_ide2024} already argued that combining programming education with the usage of IDEs helps to familiarize students with industrial technologies. Building on top of this idea, our concept is envisioned to be developed as an IDE plugin.

\paragraph{Assisted Debugging}
\label{sec:support-student}
\begin{figure}[t]
\centering
\includegraphics[width=\columnwidth]{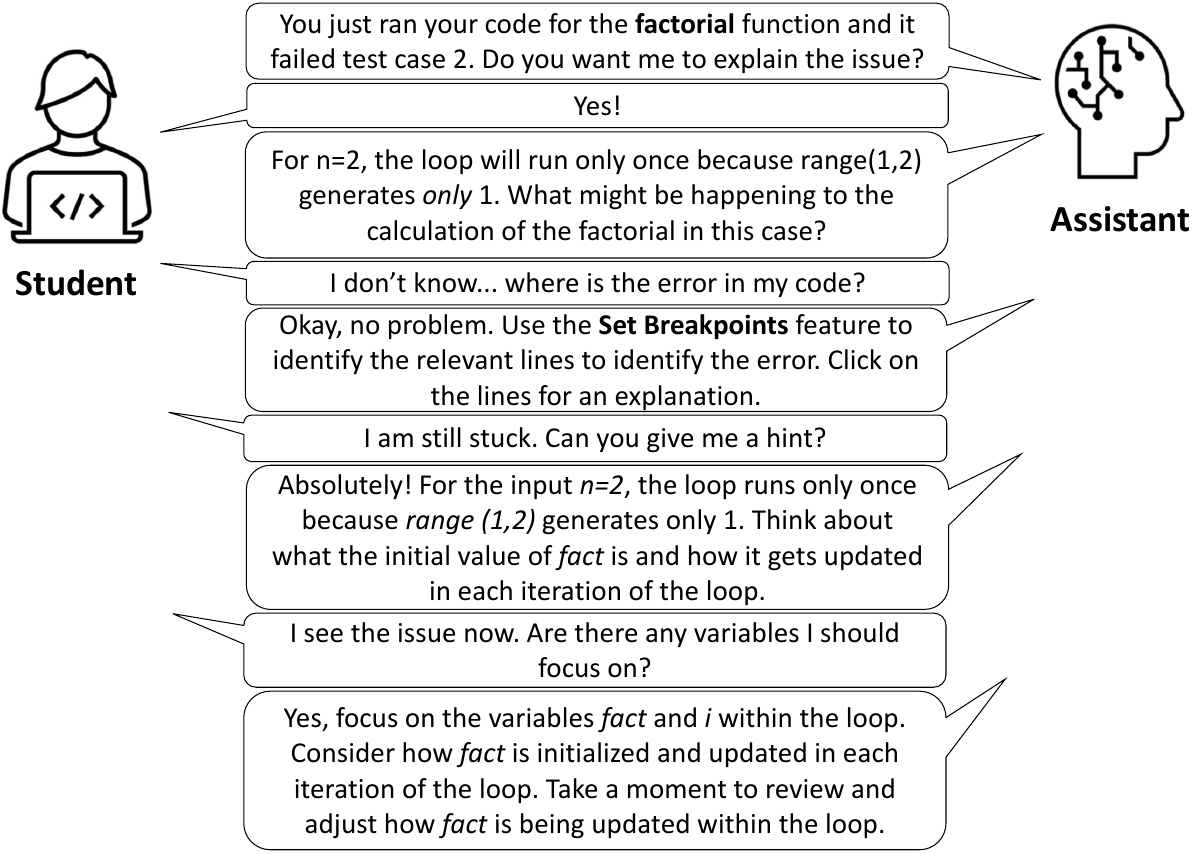}
\caption{Envisioned conversation between student and the AI assistant.}
\label{fig:conversation}
\end{figure}
Figure~\ref{fig:conversation} shows the envisioned interaction between student and AI assistant. First, the AI assistant can (1) explain failing tests and the observed failures. Next, it can (2) automatically set breakpoints at locations of interest. These locations can be identified, e.g., by using SBFL. Setting such breakpoints enables the unexperienced student to use the interactive debugger to step through the test execution.
Based on LLMs, we also can (3) generate hints in natural language explaining these breakpoints. In cases where a student's understanding of the problem is far from what is expected, the AI assistant can (4) deliver incremental guidance, e.g., via progressive hints. %, to ensure that the student understands the problem correctly.
If the student is still stuck, the AI assistant can (5) partially reveal why a certain test fails.
Furthermore, we can (6) help the student by highlighting interesting variables that the student can observe during debugging.
This interaction aims to guide the student to repair their solution and to have a proper \textit{learning} experience. % instead of just revealing the patch.

% \begin{figure*}[h]
% \centering
% \includegraphics[width=0.7\textwidth]{screenshot_cut.pdf}
% \caption{Screenshot from our prototypical implementation. The left part shows the code and the automatically set breakpoints. The right part shows the feature selection and the dialogue with the AI-based chatbot.}
% \label{fig:screenshot}
% \end{figure*}

% \begin{figure}[t]
% \centering
% \includegraphics[width=0.75\columnwidth]{experience_cut.pdf}
% \caption{Participants' programming/debugging experience.}
% \label{fig:experience}
% \end{figure}

\section{Pilot User Study}
% To provide first insights towards our research goals, we executed a pilot user study. We investigate the practicality of our assisted debugging concept, as well as the usability of our implementation.

% \begin{tcolorbox}[boxrule=1pt,left=1pt,right=1pt,top=1pt,bottom=1pt]
% Our replication package with our prototype implementation and the study artifacts are available: \url{https://figshare.com/s/489b8da27b4e0fcedbce}
% \end{tcolorbox}

\subsection{User Study Setup}

\paragraph{Prototype}
We realize our study prototype as a VS Code IDE extension.
% As a study prototype, we realized the Simulated Interactive Debugging concept as a Visual Studio Code IDE extension.
Our concept is language-agnostic; however, due to the programming courses at our institution, we decided to first focus on supporting Python. It provides two core features for the assisted debugger: the automatic setting of breakpoints (based on SBFL with FauxPy~\cite{rezaalipour2024fauxpy}) and a chat interface providing tailored debugging hints.
% Each feature can be triggered via a VSCode command.
The chat interface connects the participant with an LLM-powered chatbot using OpenAI gpt-3.5-turbo.
% The user can choose from two modes: the \textit{Generate Hints} mode, which only allows the student to get step-by-step hints, and the \textit{Interactive Debugging Guidance} mode, which allows the student to engage in a dialogue with the LLM in a controlled way.
% The student can discuss code issues and query about how to proceed with debugging or improving the code. 
% Figure~\ref{fig:screenshot} shows an annotated screenshot of our implementation.
% As underlying LLM, we used OpenAI gpt-3.5-turbo.
Note that in the implementation of our current prototype, we focused only on the debugging use case.
% omitted the support for the lecturer's and the student's test assessment.
% because our primary interest was in the feedback for the actual debugging.

\begin{figure}[t]
\centering
\includegraphics[width=\columnwidth]{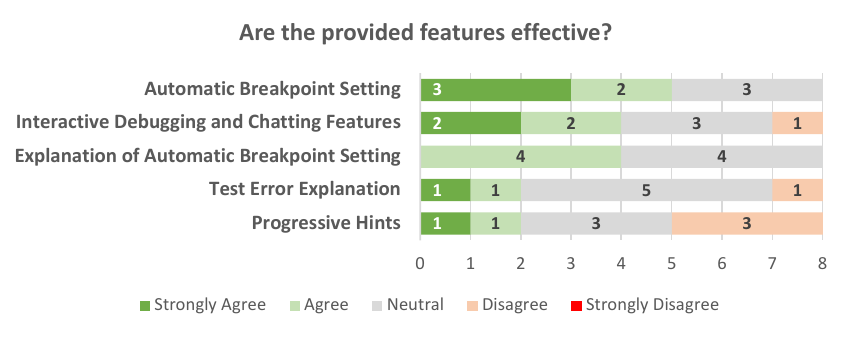}
\caption{Participants' assessment of the provided features.}
\label{fig:features}
\end{figure}

\paragraph{Experiment Structure}
The study started with a short briefing, the IRB signing, and the completion of a profiling survey. Then, we provided a 15-minute introduction into debugging and for demonstrating the tool's features. Afterwards, the participants had 40 minutes to solve two coding tasks. 
The tasks originate from LeetCode and are of medium difficulty that should not overwhelm or underwhelm the students. 
Both were provided with pre-existing buggy solutions, along with corresponding passing and failing tests.
In the first task, named \textit{maxOperations}~\cite{leetcode_maxOperations}, the participants have to fix an algorithm that is supposed to calculate the maximum number of times to pick two numbers from a given array whose sum equals the parameter $k$. In the second task, named \textit{longestOnes}~\cite{leetcode_longestOnes}, the participants have to fix the algorithm that is supposed to compute the maximum number of consecutive 1's in a binary array if one can flip at most $k$ 0's. The participants first had to understand the task's description, find the errors in the provided solution, and finally modify the program to pass all tests. We provided a handout with a possible workflow to solve the tasks with the available assisted debugging features (see artifact). We informed the participants that the generated hints might not be entirely accurate. In the end, the participants were asked to complete a usability questionnaire~\cite{brooke1996sus} and a post-task survey. The study was timed for 70 minutes, and the participants were compensated with 30 SGD.

\paragraph{Participants}
For our pilot study, we recruited eight first-year undergraduate CS students. Based on our profiling survey results, all of them are beginners in the field of programming, and hence, are in our target group:
% Figure~\ref{fig:experience} illustrates the self-classified experience levels of our participants:
most of them have programming experience of less than two years and can only write basic Python programs. 6/8 participants mentioned that they can debug basic errors but struggle with complex ones. At the same time, 6/8 have never heard of or never used an interactive debugger. Those who understand the concept of interactive debuggers still prefer using print/log statements.

\subsection{Results \& Discussion}
% In the following, we discuss the findings from our user study regarding the feedback on the \textit{existing features} of the assisted debugger, the overall \textit{usability} of our prototype, the participant's confidence of \textit{manually} setting breakpoints, and additionally desired features.
% In the following, we discuss our findings regarding the feedback on the \textit{existing features} of the assisted debugger, the overall \textit{usability}, the participant's confidence of \textit{manually} setting breakpoints, and additionally desired features.

\paragraph{Existing Features}
In total, all eight participants were able to solve the two programming tasks in the provided time.
%
%
% Overview of overall feature effectiveness ranking
As shown in Figure~\ref{fig:features}, the most effective feature is the automatic breakpoint setting (5/8 agreement), followed by interactive debugging and chatting (4/8) and the explanation of automatic breakpoints (4/8). The features for progressive hints and test error explanation are only positively evaluated by 2/8 students.
%
%Progressive Hints
The \textit{progressive hints} helped the students to identify where their code has a fault but did not reveal the solution. Instead, it tries to guide the students to find the solution on their own to foster a learning effect. However, the participants perceived this feature as often too vague or too general.
%
%Test Error explanation
For most participants, the \textit{test error explanation} helped to analyze the coding issue, while the feature itself could be made more interactively, e.g., allowing for different/alternative explanations.
%
% Explanation of Automatic Breakpoint Setting
The \textit{explanation of the automatically set breakpoints} helped most students to understand their faults. Therefore, we conclude that it is a good addition to the automatic setting of breakpoints.
%
% Interactive Debugging and Chatting Features
The \textit{interactive debugging and chatting} was able to add more detailed information to the debugging efforts. One participant mentioned that this even accelerated the overall debugging process. Others still found that the responses were vague and that more prompts could improve the feature.
%
% Automatic Breakpoint Setting
The \textit{automatic setting breakpoints} feature was considered very effective as it set the breakpoints at the right places.
One participant summarized its effectiveness as follows:
\begin{tcolorbox}[boxrule=1pt,left=1pt,right=1pt,top=1pt,bottom=1pt]
\begin{center}
\small
\textit{``It did get the breakpoint in the right place, which was a good start in debugging.''}
\end{center}
\end{tcolorbox}
One of the first hurdles of conventional interactive debugging is to set breakpoints. Our concept of \textit{simulated} interactive debugging helps the students to get started, and therefore, guides them along a deeper understanding of their coding errors and strives to develop debugging skills.
%
% Quote about AI usage.
The following quote from one of our participants about the \textit{interactive debugging and chatting} highlights another aspect:
\begin{tcolorbox}[boxrule=1pt,left=1pt,right=1pt,top=1pt,bottom=1pt]
\begin{center}
\small
\textit{``It helped me brainstorm while guiding me and \\not letting me fully rely on A''}
\end{center}
\end{tcolorbox}
Our goal is to guide students to learn debugging but also the usage of AI-based tools like ChatGPT. By constructing an LLM-based chatbot, we can control the prompts and the content of responses, and hence, still foster a learning experience for the students, e.g., by filtering direct solutions.

\paragraph{Usability}
Using the standardized system usability survey by Brooke~\cite{brooke1996sus}, we can conclude that overall, the participants are satisfied with the prototype's usability. The average System Usability Scale (SUS) is around 65 (out of 100), providing a good basis for our future work. For example, 7/8 participants would like to use the system frequently, and 5/8 agreed that the system is easy to use. % These findings confirm that our efforts are heading in the right direction leaving space for further improvement.
% C	65.0 – 71.0	41 – 59		Marginal	Passive
% -> Marginal acceptable, Comment Yannic: for a research prototype that is acutally a very good score, I would say...
% https://measuringu.com/interpret-sus-score/#:~:text=The%20average%20score%20(at%20the,the%20scores%20in%20the%20database).

\paragraph{Debugging Process}
After completing the tasks, we also asked the participants how confident they are about setting breakpoints on their own.
The responses indicated mixed confidence with a tendency to lower confidence: 1/8 participants strongly agreed to be confident, 2/8 agreed, 2/8 noted a neutral opinion, and 3/8 disagreed.
% As shown in Figure~\ref{fig:confidence}, the participants have mixed confidence about setting their own breakpoints, with a tendency to lower confidence.
% %
% \begin{figure}[h]
% \centering
% \includegraphics[width=0.9\columnwidth]{confidence.pdf}
% \caption{Participants' confidence about setting their own breakpoints.}
% \label{fig:confidence}
% \end{figure}
% %
% Afterwards, we asked them to set breakpoints for a given code snippet (see Listing~\ref{lst:manual-breakpoint}). The interesting lines are 6 to 7 because the handling of \texttt{None} values is missing.
Afterwards, we asked them to set breakpoints for a given buggy code snippet that calculates the average for a given list of grades without catering for \texttt{None} values in the list.
% %
% % basicstyle=\tiny, %or \small or \footnotesize etc.
% \begin{lstlisting}[language=Python, caption=Code snippet for setting a manual breakpoints., label=lst:manual-breakpoint, frame=single, numbers=left, numbersep=8pt, xleftmargin=20pt, xrightmargin=10pt, basicstyle=\footnotesize]
% # Calculate the average of a list of
% # grades, if None, you should continue
% # to the next grade
% def calculate_average (grades):
%     total = 0
%     for grade in grades:
%         total += grade
%     average = total / len (grades)
%     return average
% 
% # Test the function
% grades = [85, 90, 78, None, 92, None]
% average_grade = calculate_average(grades)
% print(f"The average is: {average_grade}")
% \end{lstlisting}
% %
% 6/8 participants would set breakpoints either at lines 6 or 7, or both, which shows that actually most of the participants have a correct intuition about where to set the breakpoint for further inspection. 
6/8 participants would set breakpoints at the correct lines, which shows that actually most of the participants have a correct intuition about where to set the breakpoint for further inspection. 
While we see good results for this relatively simple example, we conclude that due to overall low confidence, the automated setting of breakpoints would still be an interesting feature. However, such a feature could be applied only as hints or as confirmation for the students' manually set breakpoints.
%
% Key Differences to usual debugging
Additionally, we asked the participants about their perceived key difference to their usual debugging process. The majority (6/8) mentioned that our Simulated Interactive Debugging tool helped them to debug the coding issue and guided them systematically.
% Two participants indicated that they already use LLM-based tools like ChatGPT.
However, one of the participants mentioned that the interactive debugging process could be more time-consuming, though.
The following two quotes are representative responses:
\begin{tcolorbox}[boxrule=1pt,left=1pt,right=1pt,top=1pt,bottom=1pt]
\begin{center}
\small
\textit{``Normally I would just trial and error until I got it right, but this tool allowed me to systematically identify the problems in the code step by step.''}

\vspace{2mm}

\textit{``My usual debugging process includes using print statements and doing everything myself but sometimes I just get tired and ask ChatGPT. Simulated Interactive Debugging helped me think while guiding me through the right process.''}
\end{center}
\end{tcolorbox}

\paragraph{Additional Features}
At the end of the post-task survey, we asked the participants about any additional features they would like to see in the Simulated Interactive Debugger.
6/8 participants indicated that further hints about what and how to change would be helpful. For example, when a student really struggles, the chatbot could provide more detailed explanations and propose changes that the student could select.
%
% * also tells you what to change 
% * giving you further elaboration on how you should change a particular code when using the generative hints
% * maybe more guidance on the errors
% * MAYBE it can give us options, like which one do u think is the answer or fits the logic of the code better.(if we r really struggling haha)
% * maybe can fix the Ai chat bot to make the explanation more detailed
% * more prompts
%
One participant mentioned that the UI could be improved to allow easier switching between the assisted debugger and the interactive debugging tool in the IDE.
%
% * A cleaner UI to make it easier to switch between debugging information and the debugger tool.
%
Lastly, one participant highlighted that automated test generation would be helpful, e.g., if students cannot find an appropriate input for debugging.
%
% * perhaps a test case generation feature (mostly cause i am lazy to write my own XD)

\subsection{Threats to Validity}
We performed a first, small pilot study with eight CS students, i.e., our results may not generalize. We focused on a small number of students to receive focused feedback for the next development phase. The results of our profiling survey showed that the participants are in our target, i.e., beginners in programming with no or very limited (interactive) debugging experience. For the next cycle of user studies, we will try to reach a larger set of participants. To mitigate the threat of manual errors in our analysis, we ensured that all results were analyzed and agreed upon by two authors independently.

\section{Conclusion and Future Work}
% Conclusion
We proposed the concept of Simulated Interactive Debugging as a first step to automate the teaching of debugging skills.
% Impact
Incorporating such techniques in the CS curriculum will be essential to teach debugging beyond ad-hoc approaches and guide students in using AI-based programming tools.
% Future Work
% As part of our design science methodology, we will use the feedback from our user study to refine our concept and implementation.
In future, we plan to incorporate more active guidance, e.g., by using a state machine-driven approach, similar to Bouzenia et al.~\cite{bouzenia2024repairagent}. We will also integrate state-of-the-art APR techniques and implement the test assessment and generation features.

\section*{Data Availability}
Our artifact with the prototype and the study artifacts is available: \url{https://doi.org/10.6084/m9.figshare.28202336}
% \begin{center}
% \end{center}

\section*{Acknowledgment}
Part of this work was supported by the Ministry of Education, Singapore, under the Tertiary Research Fund (MOE-TRF) Grant No. MOE2023-TRF-034. We gratefully acknowledge this support, which made this research possible.

\bibliographystyle{IEEEtran}
\bibliography{references}

\begin{thebibliography}{10}
\providecommand{\url}[1]{#1}
\csname url@samestyle\endcsname
\providecommand{\newblock}{\relax}
\providecommand{\bibinfo}[2]{#2}
\providecommand{\BIBentrySTDinterwordspacing}{\spaceskip=0pt\relax}
\providecommand{\BIBentryALTinterwordstretchfactor}{4}
\providecommand{\BIBentryALTinterwordspacing}{\spaceskip=\fontdimen2\font plus
\BIBentryALTinterwordstretchfactor\fontdimen3\font minus \fontdimen4\font\relax}
\providecommand{\BIBforeignlanguage}[2]{{%
\expandafter\ifx\csname l@#1\endcsname\relax
\typeout{** WARNING: IEEEtran.bst: No hyphenation pattern has been}%
\typeout{** loaded for the language `#1'. Using the pattern for}%
\typeout{** the default language instead.}%
\else
\language=\csname l@#1\endcsname
\fi
#2}}
\providecommand{\BIBdecl}{\relax}
\BIBdecl

\bibitem{radermacher_icse2014}
\BIBentryALTinterwordspacing
A.~Radermacher, G.~Walia, and D.~Knudson, ``Investigating the skill gap between graduating students and industry expectations,'' in \emph{Companion Proceedings of the 36th International Conference on Software Engineering}, ser. ICSE Companion 2014.\hskip 1em plus 0.5em minus 0.4em\relax New York, NY, USA: Association for Computing Machinery, 2014, p. 291–300. [Online]. Available: \url{https://doi.org/10.1145/2591062.2591159}
\BIBentrySTDinterwordspacing

\bibitem{Michaeli2019}
\BIBentryALTinterwordspacing
T.~Michaeli and R.~Romeike, ``Improving debugging skills in the classroom: The effects of teaching a systematic debugging process,'' in \emph{Proceedings of the 14th Workshop in Primary and Secondary Computing Education}, ser. WiPSCE '19.\hskip 1em plus 0.5em minus 0.4em\relax New York, NY, USA: Association for Computing Machinery, 2019. [Online]. Available: \url{https://doi.org/10.1145/3361721.3361724}
\BIBentrySTDinterwordspacing

\bibitem{clara_pldi2018}
\BIBentryALTinterwordspacing
S.~Gulwani, I.~Radi\v{c}ek, and F.~Zuleger, ``Automated clustering and program repair for introductory programming assignments,'' in \emph{Proceedings of the 39th ACM SIGPLAN Conference on Programming Language Design and Implementation}, ser. PLDI 2018.\hskip 1em plus 0.5em minus 0.4em\relax New York, NY, USA: Association for Computing Machinery, 2018, p. 465–480. [Online]. Available: \url{https://doi.org/10.1145/3192366.3192387}
\BIBentrySTDinterwordspacing

\bibitem{refactory_ase2019}
Y.~Hu, U.~Z. Ahmed, S.~Mechtaev, B.~Leong, and A.~Roychoudhury, ``Re-factoring based program repair applied to programming assignments,'' in \emph{2019 34th IEEE/ACM International Conference on Automated Software Engineering (ASE)}, 2019, pp. 388--398.

\bibitem{pydex_oopsla2024}
\BIBentryALTinterwordspacing
J.~Zhang, J.~P. Cambronero, S.~Gulwani, V.~Le, R.~Piskac, G.~Soares, and G.~Verbruggen, ``{PyDex}: Repairing bugs in introductory python assignments using {LLMs},'' \emph{Proc. ACM Program. Lang.}, vol.~8, no. OOPSLA1, Apr. 2024. [Online]. Available: \url{https://doi.org/10.1145/3649850}
\BIBentrySTDinterwordspacing

\bibitem{fan_issta2023}
\BIBentryALTinterwordspacing
Z.~Fan, S.~H. Tan, and A.~Roychoudhury, ``Concept-based automated grading of {CS-1} programming assignments,'' in \emph{Proceedings of the 32nd ACM SIGSOFT International Symposium on Software Testing and Analysis}, ser. ISSTA 2023.\hskip 1em plus 0.5em minus 0.4em\relax New York, NY, USA: Association for Computing Machinery, 2023, p. 199–210. [Online]. Available: \url{https://doi.org/10.1145/3597926.3598049}
\BIBentrySTDinterwordspacing

\bibitem{codeaid_chi2024}
\BIBentryALTinterwordspacing
M.~Kazemitabaar, R.~Ye, X.~Wang, A.~Z. Henley, P.~Denny, M.~Craig, and T.~Grossman, ``{CodeAid}: Evaluating a classroom deployment of an {LLM}-based programming assistant that balances student and educator needs,'' in \emph{Proceedings of the 2024 CHI Conference on Human Factors in Computing Systems}, ser. CHI '24.\hskip 1em plus 0.5em minus 0.4em\relax New York, NY, USA: Association for Computing Machinery, 2024. [Online]. Available: \url{https://doi.org/10.1145/3613904.3642773}
\BIBentrySTDinterwordspacing

\bibitem{fan2024softwareengineeringeducationalexperience}
Z.~Fan, Y.~Noller, A.~Dandekar, and A.~Roychoudhury, ``Software engineering educational experience in building an intelligent tutoring system,'' in \emph{CSEE{\&}T}.\hskip 1em plus 0.5em minus 0.4em\relax {IEEE}, 2025, pp. 75--86.

\bibitem{hevner2005_designscience}
\BIBentryALTinterwordspacing
A.~R. Hevner, S.~T. March, J.~Park, and S.~Ram, ``Design science in information systems research,'' \emph{MIS Quarterly}, vol.~28, no.~1, pp. 75--105, 2004. [Online]. Available: \url{http://www.jstor.org/stable/25148625}
\BIBentrySTDinterwordspacing

\bibitem{wieringa2014_designscience}
\BIBentryALTinterwordspacing
R.~J. Wieringa, \emph{Design Science Methodology for Information Systems and Software Engineering}.\hskip 1em plus 0.5em minus 0.4em\relax Berlin, Heidelberg: Springer Berlin Heidelberg, 2014. [Online]. Available: \url{https://doi.org/10.1007/978-3-662-43839-8_1}
\BIBentrySTDinterwordspacing

\bibitem{masters2011}
K.~Masters, ``A brief guide to understanding {MOOCs},'' \emph{The Internet Journal of Medical Education}, vol.~1, no.~2, p.~2, 2011.

\bibitem{birillo_sigcse2022}
\BIBentryALTinterwordspacing
A.~Birillo, I.~Vlasov, A.~Burylov, V.~Selishchev, A.~Goncharov, E.~Tikhomirova, N.~Vyahhi, and T.~Bryksin, ``Hyperstyle: A tool for assessing the code quality of solutions to programming assignments,'' in \emph{Proceedings of the 53rd ACM Technical Symposium on Computer Science Education - Volume 1}, ser. SIGCSE 2022.\hskip 1em plus 0.5em minus 0.4em\relax New York, NY, USA: Association for Computing Machinery, 2022, p. 307–313. [Online]. Available: \url{https://doi.org/10.1145/3478431.3499294}
\BIBentrySTDinterwordspacing

\bibitem{codehelp_calling2023}
\BIBentryALTinterwordspacing
M.~Liffiton, B.~E. Sheese, J.~Savelka, and P.~Denny, ``{CodeHelp}: Using large language models with guardrails for scalable support in programming classes,'' in \emph{Proceedings of the 23rd Koli Calling International Conference on Computing Education Research}, ser. Koli Calling '23.\hskip 1em plus 0.5em minus 0.4em\relax New York, NY, USA: Association for Computing Machinery, 2024. [Online]. Available: \url{https://doi.org/10.1145/3631802.3631830}
\BIBentrySTDinterwordspacing

\bibitem{hou_ls2024}
\BIBentryALTinterwordspacing
X.~Hou, Z.~Wu, X.~Wang, and B.~J. Ericson, ``{CodeTailor}: {LLM}-powered personalized parsons puzzles for engaging support while learning programming,'' in \emph{Proceedings of the Eleventh ACM Conference on Learning @ Scale}, ser. L@S '24.\hskip 1em plus 0.5em minus 0.4em\relax New York, NY, USA: Association for Computing Machinery, 2024, p. 51–62. [Online]. Available: \url{https://doi.org/10.1145/3657604.3662032}
\BIBentrySTDinterwordspacing

\bibitem{zhao2024peer}
Q.~Zhao, F.~Liu, L.~Zhang, Y.~Liu, Z.~Yan, Z.~Chen, Y.~Zhou, J.~Jiang, and G.~Li, ``Peer-aided repairer: Empowering large language models to repair advanced student assignments,'' \emph{arXiv preprint arXiv:2404.01754}, 2024.

\bibitem{kumar_llm4code2024}
\BIBentryALTinterwordspacing
S.~S~Kumar, M.~Adam~Lones, M.~Maarek, and H.~Zantout, ``Investigating the proficiency of large language models in formative feedback generation for student programmers,'' in \emph{Proceedings of the 1st International Workshop on Large Language Models for Code}, ser. LLM4Code '24.\hskip 1em plus 0.5em minus 0.4em\relax New York, NY, USA: Association for Computing Machinery, 2024, p. 88–93. [Online]. Available: \url{https://doi.org/10.1145/3643795.3648380}
\BIBentrySTDinterwordspacing

\bibitem{abolnejadian_chiea2024}
\BIBentryALTinterwordspacing
M.~Abolnejadian, S.~Alipour, and K.~Taeb, ``Leveraging {ChatGPT} for adaptive learning through personalized prompt-based instruction: A {CS1} education case study,'' in \emph{Extended Abstracts of the CHI Conference on Human Factors in Computing Systems}, ser. CHI EA '24.\hskip 1em plus 0.5em minus 0.4em\relax New York, NY, USA: Association for Computing Machinery, 2024. [Online]. Available: \url{https://doi.org/10.1145/3613905.3637148}
\BIBentrySTDinterwordspacing

\bibitem{kurniawan_tale2023}
O.~Kurniawan, C.~M. Poskitt, I.~Al~Hoque, N.~T.~S. Lee, C.~Jégourel, and N.~Sockalingam, ``How helpful do novice programmers find the feedback of an automated repair tool?'' in \emph{2023 IEEE International Conference on Teaching, Assessment and Learning for Engineering (TALE)}, 2023, pp. 1--6.

\bibitem{birillo_koli2024}
\BIBentryALTinterwordspacing
A.~Birillo, E.~Artser, A.~Potriasaeva, I.~Vlasov, K.~Dzialets, Y.~Golubev, I.~Gerasimov, H.~Keuning, and T.~Bryksin, ``One step at a time: Combining {LLMs} and static analysis to generate next-step hints for programming tasks,'' in \emph{Proceedings of the 24th Koli Calling International Conference on Computing Education Research}, ser. Koli Calling '24.\hskip 1em plus 0.5em minus 0.4em\relax New York, NY, USA: Association for Computing Machinery, 2024. [Online]. Available: \url{https://doi.org/10.1145/3699538.3699556}
\BIBentrySTDinterwordspacing

\bibitem{koutcheme2024benchmarking}
C.~Koutcheme, N.~Dainese, S.~Sarsa, J.~Leinonen, A.~Hellas, and P.~Denny, ``Benchmarking educational program repair,'' \emph{arXiv preprint arXiv:2405.05347}, 2024.

\bibitem{phung_icer2023}
\BIBentryALTinterwordspacing
T.~Phung, V.-A. P\u{a}durean, J.~Cambronero, S.~Gulwani, T.~Kohn, R.~Majumdar, A.~Singla, and G.~Soares, ``Generative {AI} for programming education: Benchmarking {ChatGPT}, {GPT-4}, and human tutors,'' in \emph{Proceedings of the 2023 ACM Conference on International Computing Education Research - Volume 2}, ser. ICER '23.\hskip 1em plus 0.5em minus 0.4em\relax New York, NY, USA: Association for Computing Machinery, 2023, p. 41–42. [Online]. Available: \url{https://doi.org/10.1145/3568812.3603476}
\BIBentrySTDinterwordspacing

\bibitem{McCauley01062008}
R.~McCauley, S.~Fitzgerald, G.~Lewandowski, L.~Murphy, B.~Simon, L.~Thomas, and C.~Zander, ``Debugging: a review of the literature from an educational perspective,'' \emph{Computer Science Education}, vol.~18, no.~2, pp. 67--92, 2008.

\bibitem{alhossami_sigcse2024}
\BIBentryALTinterwordspacing
E.~Al-Hossami, R.~Bunescu, J.~Smith, and R.~Teehan, ``Can language models employ the {Socratic} method? {Experiments} with code debugging,'' in \emph{Proceedings of the 55th ACM Technical Symposium on Computer Science Education V. 1}, ser. SIGCSE 2024.\hskip 1em plus 0.5em minus 0.4em\relax New York, NY, USA: Association for Computing Machinery, 2024, p. 53–59. [Online]. Available: \url{https://doi.org/10.1145/3626252.3630799}
\BIBentrySTDinterwordspacing

\bibitem{padurean2024bugspotter}
\BIBentryALTinterwordspacing
V.-A. Padurean, P.~Denny, and A.~Singla, ``Bugspotter: Automated generation of code debugging exercises,'' in \emph{Proceedings of the 56th ACM Technical Symposium on Computer Science Education V. 1}, ser. SIGCSETS 2025.\hskip 1em plus 0.5em minus 0.4em\relax New York, NY, USA: Association for Computing Machinery, 2025, p. 896–902. [Online]. Available: \url{https://doi.org/10.1145/3641554.3701974}
\BIBentrySTDinterwordspacing

\bibitem{shein_cacm2024}
\BIBentryALTinterwordspacing
E.~Shein, ``The impact of {AI} on computer science education,'' \emph{Commun. ACM}, vol.~67, no.~9, p. 13–15, Aug. 2024. [Online]. Available: \url{https://doi.org/10.1145/3673428}
\BIBentrySTDinterwordspacing

\bibitem{denny_iticse2024}
\BIBentryALTinterwordspacing
P.~Denny, S.~MacNeil, J.~Savelka, L.~Porter, and A.~Luxton-Reilly, ``Desirable characteristics for {AI} teaching assistants in programming education,'' in \emph{Proceedings of the 2024 on Innovation and Technology in Computer Science Education V. 1}, ser. ITiCSE 2024.\hskip 1em plus 0.5em minus 0.4em\relax New York, NY, USA: Association for Computing Machinery, 2024, p. 408–414. [Online]. Available: \url{https://doi.org/10.1145/3649217.3653574}
\BIBentrySTDinterwordspacing

\bibitem{kazemitabaar2024exploring}
M.~Kazemitabaar, O.~Huang, S.~Suh, A.~Z. Henley, and T.~Grossman, ``Exploring the design space of cognitive engagement techniques with {AI}-generated code for enhanced learning,'' in \emph{{IUI}}.\hskip 1em plus 0.5em minus 0.4em\relax {ACM}, 2025, pp. 695--714.

\bibitem{birillo_ide2024}
\BIBentryALTinterwordspacing
A.~Birillo, M.~Tigina, Z.~Kurbatova, A.~Potriasaeva, I.~Vlasov, V.~Ovchinnikov, and I.~Gerasimov, ``Bridging education and development: Ides as interactive learning platforms,'' in \emph{Proceedings of the 1st ACM/IEEE Workshop on Integrated Development Environments}, ser. IDE '24.\hskip 1em plus 0.5em minus 0.4em\relax New York, NY, USA: Association for Computing Machinery, 2024, p. 53–58. [Online]. Available: \url{https://doi.org/10.1145/3643796.3648454}
\BIBentrySTDinterwordspacing

\bibitem{fl_survey_tse2016}
W.~E. Wong, R.~Gao, Y.~Li, R.~Abreu, and F.~Wotawa, ``A survey on software fault localization,'' \emph{IEEE Transactions on Software Engineering}, vol.~42, no.~8, pp. 707--740, 2016.

\bibitem{rezaalipour2024fauxpy}
M.~Rezaalipour and C.~A. Furia, ``{FauxPy}: A fault localization tool for {Python},'' \emph{arXiv preprint arXiv:2404.18596}, 2024.

\bibitem{leetcode_maxOperations}
``{LeetCode:} max number of k-sum pairs,'' \url{https://leetcode.com/problems/max-number-of-k-sum-pairs}, 2025, accessed: 2025-01-10.

\bibitem{leetcode_longestOnes}
``{LeetCode:} max consecutive ones {III},'' \url{https://leetcode.com/problems/max-consecutive-ones-iii}, 2025, accessed: 2025-01-10.

\bibitem{brooke1996sus}
J.~Brooke, ``{SUS}: A quick and dirty usability scale,'' \emph{Usability Evaluation in Industry}, 1996.

\bibitem{bouzenia2024repairagent}
I.~Bouzenia, P.~T. Devanbu, and M.~Pradel, ``{RepairAgent}: An autonomous, {LLM}-based agent for program repair,'' in \emph{{ICSE}}.\hskip 1em plus 0.5em minus 0.4em\relax {IEEE}, 2025, pp. 2188--2200.

\end{thebibliography}

\end{document}